\documentclass[11pt, a4paper]{article}
\usepackage[T1]{fontenc}
\usepackage{amsfonts}
\usepackage{amsmath}
\usepackage{amssymb}
\usepackage{amsthm}
\usepackage{bbm}
\usepackage{bm}
\usepackage{mathrsfs}
\usepackage{verbatim}
\usepackage{setspace}
\usepackage{enumitem}

\theoremstyle{plain}
\newtheorem{Thm}{Theorem}[section]
\newtheorem{Prop}[Thm]{Proposition}
\newtheorem{Lemma}[Thm]{Lemma}
\newtheorem{Cor}[Thm]{Corollary}
\theoremstyle{definition}

\newtheorem{Remark}[Thm]{Remark}
\newtheorem{Fact}[Thm]{Fact}
\newtheorem{Example}[Thm]{Example}
\theoremstyle{remark}

\renewenvironment{thebibliography}[1]{%
\begin{oldthebibliography}{#1}%
\setlength{\baselineskip}{.9em}
\small
\setlength{\parskip}{0ex}%
\setlength{\itemsep}{0.1em}%
}%
{%
\end{oldthebibliography}%
}
\newcommand{\q}{\quad}

\newcommand{\R}{\mathbb{R}}

\newcommand{\cF}{\mathcal{F}}

\newcommand{\cA}{\mathcal{A}}

\newcommand{\cE}{\mathcal{E}}

\newcommand{\sC}{\mathscr{C}}
\newcommand{\sD}{\mathscr{D}}

\newcommand{\sX}{\mathscr{X}}
\newcommand{\sM}{\mathscr{M}}

\newcommand{\sY}{\mathscr{Y}}

\newcommand{\br}[1]{\langle #1 \rangle}

\DeclareMathOperator{\sgn}{sign}

\DeclareMathOperator{\esssup}{ess\, sup}
\DeclareMathOperator{\essinf}{ess\, inf}

\DeclareMathOperator{\argmax}{arg\, max}
\newcommand{\bmat}{\begin{pmatrix}}
\newcommand{\emat}{\end{pmatrix}}

\newcommand{\hX}{\widehat{X}}
\newcommand{\hY}{\widehat{Y}}

\newcommand{\tc}{\tilde{c}}
\newcommand{\hc}{\hat{c}}
\newcommand{\cpi}{\check{\pi}}
\newcommand{\fg}{\mathfrak{g}}
\newcommand{\cc}{\check{c}}
\newcommand{\hpi}{\hat{\pi}}
\newcommand{\hkappa}{\hat{\kappa}}
\newcommand{\tpi}{\tilde{\pi}}

\newcommand{\tY}{\widetilde{Y}}

\newcommand{\sint}{\stackrel{\mbox{\tiny$\bullet$}}{}}

\numberwithin{equation}{section}

\begin{document}%

\title{The Opportunity Process for Optimal Consumption and Investment with Power Utility\\[1em] \small{Marcel Nutz\\
ETH Zurich, Department of Mathematics, 8092 Zurich, Switzerland\\ \texttt{marcel.nutz@math.ethz.ch} \\
 First Version: November 24, 2009. This Version: May 31, 2010.}}%
\date{}
\maketitle \vspace{-1.5cm}

\begin{abstract}
  We study the utility maximization problem for power utility random fields in a semimartingale
  financial market, with and without intermediate consumption. The notion of an opportunity process is introduced as a reduced form of the value process
  of the resulting stochastic control problem. We show how the opportunity process describes the key objects: optimal strategy, value function, and dual problem. The results are applied to obtain monotonicity properties of the optimal consumption.
\end{abstract}

{\small


}\\

\noindent \emph{Acknowledgements.} Financial support by Swiss National Science Foundation Grant PDFM2-120424/1 is gratefully acknowledged. The author thanks Gordan \v{Z}itkovi\'c for a remark concerning Lemma~\ref{le:MartingaleOfDuality} and Martin Schweizer, Nicholas Westray and two anonymous referees for detailed comments on an earlier version of the manuscript.

\section{Introduction}

We consider the utility maximization problem in a semimartingale model for a %
financial market, with and without intermediate consumption. While the model is general,
we focus on \emph{power utilities}. %
If the maximization is seen as a stochastic control problem, the homogeneity of these utilities leads to a
factorization of the value process into a power of the current wealth and a process $L$ around which our analysis is built.
This corresponds to the usual factorization of the value function in a Markovian setting.
The process $L$ is called \emph{opportunity process} as $L_t$ encodes the conditional expected utility that can be attained from time $t$.
This name was introduced by \v{C}ern\'y and Kallsen~\cite{CernyKallsen.07} in the context of mean-variance hedging for an object that is analogous, although introduced in a different way by those authors. Surprisingly, there exists no general study of $L$ for the case of power utility, which is a gap we try to fill here.

The opportunity process is a suitable tool to derive qualitative results about the optimal consumption strategy. Indeed, we first establish the connection between $L$ and the solution $\hY$ of the convex-dual problem. Since $\hY$ is related to the optimal
consumption by the marginal utility, this leads to a \emph{feedback formula for the optimal consumption} in terms of $L$ for general semimartingale models. Previous results in this direction (see Stoikov and Zariphopoulou~\cite{StoikovZariphopoulou.05}) required a Markovian model and the verification of a solution of the Hamilton-Jacobi-Bellman equation.
Via the feedback formula, seemingly abstract results about the opportunity process translate to properties of the optimal consumption which are of direct economic interest. In particular, we derive monotonicity properties and bounds that are quite explicit despite the generality of the model.

The present paper combines tools from convex duality and dynamic programming to study the utility maximization problem from  a ``global'' point of view. The study of the \emph{local} structure requires a more computational approach presented in a companion paper~\cite{Nutz.09b} and yields, in particular, a formula for the optimal trading strategy in terms of the opportunity process. That formula cannot be obtained by the abstract arguments of the present paper. However, its derivation requires the structures that we introduce here, and therefore some details in our exposition are motivated by the requirements of the companion paper.

This paper is organized as follows.
After the introduction, we discuss power utility random fields and specify the optimization problem in detail. Section~\ref{se:oppProc} introduces the opportunity process $L$ via dynamic programming and examines its basic properties. Section~\ref{se:OppProcAndDuality} relates $L$ to convex duality theory and reverse H\"older inequalities, which is useful to obtain bounds for
the opportunity process.
Section~\ref{se:ApplicationsOfFeedbackFormula} contains the feedback formula and the applications to the study of the optimal consumption. Section~\ref{se:tradingStrategy} completes the picture by a brief description of the formula for the optimal trading strategy.
Two appendices supply the necessary results about dynamic programming and duality theory.

We refer to Jacod and Shiryaev~\cite{JacodShiryaev.03} for unexplained notation.

\section{The Optimization Problem}

\paragraph{Financial Market.} We fix the time horizon $T\in(0,\infty)$ and a filtered probability space $(\Omega,\cF,(\cF_t)_{t\in[0,T]},P)$  satisfying the usual assumptions of right-continuity and completeness, as well as $\cF_0=\{\emptyset,\Omega\}$ $P$-a.s.
We consider an $\R^d$-valued c\`adl\`ag semimartingale $R$ with $R_0=0$.
The (componentwise) stochastic exponential $S=\cE(R)$ represents the discounted price processes of $d$ risky assets,
while $R$ stands for their returns.
Our agent also has a bank account paying zero interest at his disposal.

\paragraph{Trading Strategies and Consumption.}
The agent is endowed with a deterministic initial capital $x_0>0$. A \emph{trading strategy} is a predictable $R$-integrable $\R^d$-valued process $\pi$, where the $i$th component is interpreted as the fraction of wealth (or the portfolio proportion) invested in the $i$th risky asset. A \emph{consumption strategy} is a nonnegative optional process $c$ such that $\int_0^Tc_t\,dt<\infty$ $P$-a.s.
We want to consider two cases. Either consumption occurs only at the terminal time $T$ (utility from ``terminal wealth'' only); or there is intermediate consumption plus a bulk consumption at the time horizon. To unify the notation, we define the measure $\mu$ on $[0,T]$ by
 \[
   \mu(dt):=
  \begin{cases}
    0 & \text{in the case without intermediate consumption}, \\
    dt & \text{in the case with intermediate consumption}.
  \end{cases}
 \]
 We also define $\mu^\circ:=\mu + \delta_{\{T\}}$, where $\delta_{\{T\}}$ is the unit Dirac measure at $T$.
The \emph{wealth process} $X(\pi,c)$ corresponding to a pair $(\pi,c)$ is described by the linear equation
\begin{equation}\label{eq:wealthDefinition}
   X_t(\pi,c)=x_0+\int_0^t  X_{s-}(\pi,c) \pi_s\,dR_s-\int_0^t c_s\,\mu(ds),\quad 0\leq t\leq T
\end{equation}
and the set of \emph{admissible} trading and consumption pairs is
\[
  \cA(x_0)=\big\{(\pi,c):\, X(\pi,c)>0,\, X_-(\pi,c)>0 \mbox{ and } c_T=X_T(\pi,c)\big\}.
\]
The convention $c_T=X_T(\pi,c)$ means that all the remaining wealth is consumed at time $T$; it is merely
for notational convenience. Indeed, $X(\pi,c)$ does not depend on $c_T$, hence any given consumption strategy
$c$ can be redefined to satisfy $c_T=X_T(\pi,c)$.
We fix the initial capital $x_0$ and usually write $\cA$ for $\cA(x_0)$.
A consumption strategy $c$ is called admissible if there exists $\pi$ such that $(\pi,c)\in\cA$; we write
$c\in\cA$ for brevity. The meaning of $\pi\in\cA$ is analogous.

Sometimes it is convenient to parametrize the consumption strategies as fractions of wealth.
Let $(\pi,c)\in\cA$ and let $X=X(\pi,c)$ be the corresponding wealth process. Then
\begin{equation}\label{eq:DefPropToConsume}
  \kappa:=\frac{c}{X}
\end{equation}
is called the \emph{propensity to consume} corresponding to $(\pi,c)$.
Note that $\kappa_T=1$ due to our convention that $c_T=X_T$.

\begin{Remark}\label{rk:PropensityToConsume}
  (i)~~The parametrization $(\pi,\kappa)$ allows to express wealth processes as stochastic exponentials: by~\eqref{eq:wealthDefinition},
  \begin{equation}\label{eq:wealthExponential}
     X(\pi,\kappa)=x_0\cE\big(\pi\sint R - \kappa \sint \mu\big)
  \end{equation}
  coincides with $X(\pi,c)$ for $\kappa:=c/X(\pi,c)$, where we have used that the c\`adl\`ag property implies $X(\pi,c)=X(\pi,c)_-$ $\mu$-a.e. The symbol $\sint$ indicates an integral, e.g., $\pi\sint R=\int\pi_s\,dR_s$.

  (ii)~~Relation~\eqref{eq:DefPropToConsume} induces a one-to-one correspondence between the pairs $(\pi,c)\in\cA$ and the pairs $(\pi,\kappa)$ such that $\pi\in\cA$ and $\kappa$ is a nonnegative optional process satisfying $\int_0^T \kappa_s\,ds<\infty$ $P$-a.s.\ and $\kappa_T=1$.
  Indeed, given $(\pi,c)\in\cA$, define $\kappa$ by~\eqref{eq:DefPropToConsume} with $X=X(\pi,c)$. As $X,X_->0$ and as $X$ is c\`adl\`ag, almost every path of $X$ is bounded away from zero and $\kappa$ has the desired integrability. Conversely, given $(\pi,\kappa)$, define
  $X$ via~\eqref{eq:wealthExponential} and $c:=\kappa X$; then $X=X(\pi,c)$. From admissibility we deduce $\pi^\top \Delta R >-1$ up to evanescence, which in turn shows $X>0$. Now $X_->0$ by a standard property of stochastic exponentials~\cite[II.8a]{JacodShiryaev.03}, so $(\pi,c)\in\cA$.
\end{Remark}

\paragraph{Preferences.} %
Let $D$ be a c\`adl\`ag adapted strictly positive process such that
$E\big[\int_0^T D_s\,\mu^\circ(ds)\big]<\infty$
and fix $p\in (-\infty,0)\cup(0,1)$. We define the utility random field
\begin{equation*}%
  U_t(x):=D_t\tfrac{1}{p}x^p,\q x\in [0,\infty),\; t\in [0,T],
\end{equation*}
where $1/0:=\infty$. We remark that Zariphopoulou~\cite{Zariphopoulou.01} and Tehranchi~\cite{Tehranchi.04} have previously used utility functions modified by certain multiplicative random variables, in the case where utility is obtained from terminal wealth. To wit, $U_t(x)$ is \emph{any $p$-homogeneous} utility random field such that a constant consumption yields finite
expected utility, and therefore the most  general utility random field that gives rise to the structure studied in this paper.
In particular, our results do not apply to the additive specification
$U'_t(x):=\tfrac{1}{p}(x+D_t)^p$ that would correspond to a hedging or random endowment problem, except of course for trivial choices of $D$. %

In the sequel, we will sometimes assume that there are constants $k_1$ and $k_2$ such that
\begin{equation}\label{eq:BoundsR}
  0<k_1\leq D_t\leq k_2,\q t\in [0,T].
\end{equation}
The \emph{expected utility} corresponding to a consumption strategy $c\in\cA$ is given by
$E[\int_0^T U_t(c_t)\,\mu^\circ(dt)]$.
We recall that this is either $E[U_T(c_T)]$ or $E[\int_0^T U_t(c_t)\,dt+U_T(c_T)]$. In the case without intermediate consumption, $U_t$ is irrelevant for $t<T$.

\begin{Remark}\label{rk:InterpretationsForD}
  The process $D$ can be used for discounting utility and consumption, or to determine the weight of intermediate consumption compared to terminal wealth.
  Our utility functional can also be related to the usual power utility function $\tfrac{1}{p}x^p$ in the following ways.
  If we write
  \[
    E\Big[\int_0^T U_t(c_t)\,\mu^\circ(dt)\Big]=E\Big[\int_0^T \tfrac{1}{p}c_t^p\,dK_t\Big]
  \]
  for $dK_t:=D_t\,\mu^\circ(dt)$, we have the usual power utility, but with a \emph{stochastic clock} $K$ (cf.\ Goll and Kallsen~\cite{GollKallsen.03}). In fact, one could also consider more general measures $dK$ and obtain a structure similar to our results below. %

  To model \emph{taxation} of the consumption, let $\varrho>-1$ be the tax rate and $D:=(1+\varrho)^{-p}$. If $c$ represents the cashflow out of the portfolio,
  $c/(1+\varrho)$ is the effectively obtained amount of the consumption good, yielding the instantaneous utility
  $\tfrac{1}{p}(c_t/(1+\varrho_t))^p=U_t(c_t)$. Similarly, $D_T$ can model a multiplicative \emph{bonus payment}.

  For yet another alternative, assume either that there is no intermediate consumption or that $D$ is a martingale, and that $E[D_T]=1$. Then
  \[
    E\Big[\int_0^T U_t(c_t)\,\mu^\circ(dt)\Big]=E^{\widetilde{P}}\Big[\int_0^T \tfrac{1}{p}c_t^p\,\mu^\circ(dt)\Big]%
  \]
  with the equivalent probability $\widetilde{P}$ defined by $d\widetilde{P}=D_T\,dP$. This is the standard power utility problem for an agent with \emph{subjective beliefs}, i.e., who uses $\widetilde{P}$ instead of the objective probability $P$.
\end{Remark}

We assume that the value of the utility maximization problem is finite:
\begin{equation}\label{eq:PrimalProblemFinite}
  u(x_0):=\sup_{c\in\cA(x_0)}E\Big[\int_0^T U_t(c_t)\,\mu^\circ(dt)\Big]<\infty.
\end{equation}
This is a \textbf{standing assumption} for the entire paper. It is void if \mbox{$p<0$} because then $U<0$. If $p>0$, it needs to be checked on a case-by-case basis (see also Remark~\ref{rk:suffCondFiniteness}).
A strategy $(\hpi,\hc)\in\cA(x_0)$
is \emph{optimal} if $E\big[\int_0^T U_t(c_t)\,\mu^\circ(dt)\big]=u(x_0)$. Of course, a no-arbitrage property is required to guarantee its existence. Let $\sM^S$ be the set of equivalent $\sigma$-martingale measures for $S$. If
\begin{equation}\label{eq:ELMMexists}
  \sM^S\neq\emptyset,
\end{equation}
arbitrage is excluded in the sense of the NFLVR condition (see Delbaen and Schachermayer~\cite{DelbaenSchachermayer.98}).
We can cite the following existence result of
Karatzas and \v{Z}itkovi\'c~\cite{KaratzasZitkovic.03}; it was previously obtained by
Kramkov and Schachermayer~\cite{KramkovSchachermayer.99} for the case without intermediate consumption.

\begin{Prop}\label{pr:ExistenceKZ}
  Under~\eqref{eq:BoundsR} and~\eqref{eq:ELMMexists}, there exists an optimal strategy $(\hpi,\hc)\in\cA$.
  The corresponding wealth process $\hX=X(\hpi,\hc)$ is unique. The consumption strategy $\hc$ can be chosen
  to be c\`adl\`ag and is unique $P\otimes\mu^\circ$-a.e.
\end{Prop}

In the sequel, $\hat{c}$ denotes a c\`adl\`ag version.
We note that under~$\eqref{eq:ELMMexists}$, the requirement $X(\pi,c)_->0$ in the definition of
$\cA$ is automatically satisfied as soon as $X(\pi,c)>0$, because $X(\pi,c)$ is then a positive supermartingale under an equivalent measure.

\begin{Remark}
  In Proposition~\ref{pr:ExistenceKZ}, the assumption on $D$ can be weakened by exploiting that~\eqref{eq:ELMMexists} is invariant under equivalent changes of measure. Suppose that $D=D'D''$, where $D'$ meets~\eqref{eq:BoundsR} and $D''$ is a martingale with unit expectation.
  As in Remark~\ref{rk:InterpretationsForD}, we consider the problem under the probability $d\widetilde{P}=D_T'' \,dP$, then Proposition~\ref{pr:ExistenceKZ} applies under $\widetilde{P}$ with $D'$ instead of $D$, and we obtain the existence of a solution also under $P$.
\end{Remark}

\section{The Opportunity Process}\label{se:oppProc}

This section introduces the main object under discussion. We do not yet impose the existence of an optimal strategy, but recall the standing assumption~\eqref{eq:PrimalProblemFinite}. To apply dynamic programming, we introduce for each $(\pi,c)\in\cA$ and $t\in[0,T]$ the set
\begin{equation}\label{eq:admissibleAftert}
  \cA(\pi,c,t)=\big\{(\tilde{\pi},\tilde{c})\in\cA:\, (\tilde{\pi},\tilde{c})=(\pi,c)\mbox{ on }[0,t]\big\}.
\end{equation}
These are the controls available on $(t,T]$ after having used $(\pi,c)$ until $t$.
The notation $\tilde{c}\in\cA(\pi,c,t)$ means that there exists $\tilde{\pi}$ such that $(\tilde{\pi},\tilde{c})\in\cA(\pi,c,t)$.
Given $(\pi,c)\in\cA$, we consider the \emph{value process}
\begin{equation}\label{eq:valueProcJ}
  J_t(\pi,c):=\mathop{\esssup}_{\tc\in\cA(\pi,c,t)} E\Big[\int_t^T U_s(\tc_s)\,\mu^\circ(ds)\Big|\cF_t\Big].
\end{equation}
We choose the c\`adl\`ag version of this process (see Proposition~\ref{pr:martOptPrinciple} in the Appendix). The $p$-homogeneity of the utility functional leads to the following factorization of $J$.

\begin{Prop}\label{pr:OppProcIndep}
  There exists a unique c\`adl\`ag semimartingale $L$, called \emph{opportunity process}, such that
  \begin{equation}\label{eq:OppProcIndep}
      L_t\,\tfrac{1}{p}\big(X_t(\pi,c)\big)^p= J_t(\pi,c)= \mathop{\esssup}_{\tc\in\cA(\pi,c,t)} E\Big[\int_t^T U_s(\tc_s)\,\mu^\circ(ds)\Big|\cF_t\Big]
  \end{equation}
  for any admissible strategy $(\pi,c)\in\cA$. In particular, $L_T=D_T$.
\end{Prop}

\begin{proof}
  Although the statement seems to be well known for several special cases of our setting, %
  we give a detailed proof in view of the importance for this paper. Let $(\pi,c),(\cpi,\cc)\in\cA$ and $X:=X(\pi,c)$, $\check{X}:=X(\cpi,\cc)$. We claim that
  \begin{align}\label{eq:proofOppProcIndep}
    \frac{1}{\check{X}_t^p} \mathop{\esssup}_{\tc\in\cA(\cpi,\cc,t)} E\Big[\int_t^T & U_s(\tc_s)\,\mu^\circ(ds)\Big|\cF_t\Big]\\[-5pt]
      & = \frac{1}{X_t^p}\mathop{\esssup}_{\tc\in\cA(\pi,c,t)} E\Big[\int_t^T U_s(\tc_s)\,\mu^\circ(ds)\Big|\cF_t\Big].\nonumber
  \end{align}
  Indeed, using the lattice property given in Fact~\ref{fa:latticeProperty}, we can find a sequence $(c^n)$ in $\cA(\cpi,\cc,t)$ such that, with a monotone increasing limit,
  \begin{align*}
      \frac{X_t^p}{\check{X}_t^p} \mathop{\esssup}_{\tc\in\cA(\cpi,\cc,t)} E\Big[\int_t^T U_s(\tc_s)\,\mu^\circ(ds)\Big|\cF_t\Big]
     &= \frac{X_t^p}{\check{X}_t^p} \lim_n E\Big[\int_t^T U_s(c_s^n)\,\mu^\circ(ds)\Big|\cF_t\Big]\\
     =  \lim_n E\Big[\int_t^T U_s\big(\tfrac{X_t}{\check{X}_t}c_s^n\big)\,\mu^\circ(ds)\Big|\cF_t\Big]
     &\leq \mathop{\esssup}_{\tc\in\cA(\pi,c,t)} E\Big[\int_t^T U_s(\tc_s)\,\mu^\circ(ds)\Big|\cF_t\Big],
  \end{align*}
  where we have used Fact~\ref{fa:stabilityProperty} in the last step. The claim follows by symmetry.
  Thus, if we define $L_t:=J_t(\pi,c)/\big[\tfrac{1}{p}\big(X_t(\pi,c)\big)^p\big]$, $L$ does not depend on the choice of $(\pi,c)\in\cA$ and inherits the properties of $J(\pi,c)$ and $X(\pi,c)>0$.
\end{proof}

The opportunity process describes ($p$ times) the maximal amount of conditional expected utility that can be accumulated on $[t,T]$ from one unit of wealth. In particular, the value function~\eqref{eq:PrimalProblemFinite} can be expressed as $u(x)=L_0 \frac{1}{p}x^p$.

In a Markovian setting, the factorization of the value function (which then replaces the value process) is very classical; for instance, it can already be found in Merton~\cite{Merton.71}. In that setting there is also a number of cases where $L$ is known explicitly for the case without intermediate consumption. See, e.g., Kraft~\cite{Kraft.05} for Heston's model and Kallsen and Muhle-Karbe~\cite{KallsenMuhleKarbe.08} for certain affine models including the CGMY model. For exponential L\'evy models an explicit solution is available also for the case with consumption (see Example~\ref{ex:levy}).

Mania and Tevzadze~\cite{ManiaTevzadze.03} study power utility from terminal wealth in a continuous semimartingale model; that paper contains some of the basic notions used here as well.
In fact, the opportunity process is present---in a more or less explicit form---in almost all works dealing with power utility. However, since it is impossible to discuss here this vast literature, we confine ourselves to indicating the most closely related references throughout this article. %

\begin{Remark}\label{rk:measureChange}
  Let $D$ be a martingale with $D_0=1$ and $\widetilde{P}$ as in Remark~\ref{rk:InterpretationsForD}. Bayes' rule and~\eqref{eq:OppProcIndep} show that $\widetilde{L}:=L/D$ can be understood as ``opportunity process under $\widetilde{P}$''
  for the standard power utility function.
\end{Remark}

\begin{Remark}\label{rk:PropensityToConsumeConditionalOptimality}
  We can now formalize the fact that the optimal strategies (in a suitable parametrization) do not depend on the current level of wealth, a special feature implied by the choice of power utility.
  If $(\hpi,\hc)\in\cA$ is optimal, $\hX=X(\hpi,\hc)$, and $\hkappa=\hc/\hX$ is the optimal propensity to consume,
  then $(\hpi,\hkappa)$ defines a conditionally optimal strategy for the problem
  \[
    \mathop{\esssup}_{\tc\in\cA(\pi,c,t)} E\Big[\int_t^T U_s(\tc_s)\,\mu^\circ(ds)\Big|\cF_t\Big];\quad\mbox{for \emph{any} } (\pi,c)\in\cA, \,t\in [0,T].
  \]
  To see this, fix $(\pi,c)\in \cA$ and $t\in [0,T]$. We define the pair $(\bar{\pi},\bar{c})$ by $\bar{\pi}=\pi 1_{[0,t]}+ \hpi 1_{(t,T]}$ and
  $\bar{c}= c 1_{[0,t]}+ \frac{X_t(\pi,c)}{\hX_t}\hc 1_{(t,T]}$ and let $\bar{X}:=X(\bar{\pi},\bar{c})$.
  Note that $(\hpi,\hc)$ is conditionally optimal in $\cA(\hpi,\hc,t)$, as otherwise Fact~\ref{fa:latticeProperty} yields a contradiction to the global optimality of $(\hpi,\hc)$. Now~\eqref{eq:proofOppProcIndep} with $(\cpi,\cc):=(\hpi,\hc)$ shows that
  $(\bar{\pi},\bar{c})$ is conditionally optimal in $\cA(\pi,c,t)$. The result follows as $\bar{c}/\bar{X}=\hc/\hX=\hkappa$ on $(t,T]$ by Fact~\ref{fa:stabilityProperty}.
\end{Remark}

The martingale optimality principle of dynamic programming takes the following form in our setting.

\begin{Prop}\label{pr:martOptPrincipleForL}
  Let $(\pi,c)\in\cA$ be an admissible strategy and assume
  that $E[\int_0^T U_s(c_s)\,\mu^\circ(ds)]>-\infty$. Then the process
  \[
    L_t\tfrac{1}{p}\big(X_t(\pi,c)\big)^p + \int_0^t U_s(c_s)\,\mu(ds),\quad t\in[0,T]
  \]
  is a supermartingale; it is a martingale if and only if $(\pi,c)$ is optimal.
\end{Prop}

\begin{proof}
  Combine Proposition~\ref{pr:OppProcIndep} and  Proposition~\ref{pr:martOptPrinciple}.
\end{proof}

The following lemma collects some elementary properties of $L$. The bounds are obtained by comparison with no-trade strategies, hence they are independent of the price process. If $D$ is deterministic or if there are constants $k_1,k_2>0$ as in~\eqref{eq:BoundsR}, we obtain bounds which are model-independent; they depend only on the utility function and the time to maturity.

\begin{Lemma}\label{le:BoundsForL}
  The opportunity process $L$ is a special semimartingale.
  \begin{enumerate}[topsep=2pt, partopsep=0pt, itemsep=1pt,parsep=2pt]
    \item If $p\in (0,1)$, $L$ is a supermartingale satisfying
    \begin{equation}\label{eq:BoundForL1}
      L_t\geq \big(\mu^\circ[t,T]\big)^{-p}\,E\Big[\int_t^T D_s\, \mu^\circ(ds)\Big|\cF_t\Big],\quad 0\leq t\leq T
    \end{equation}
    and $L,L_->0$. In particular, $L\geq k_1$ if $D\geq k_1$.
    \item If $p<0$, $L$ satisfies
    \begin{equation}\label{eq:BoundForL2}
      0\leq L_t\leq \big(\mu^\circ[t,T]\big)^{-p}\,E\Big[\int_t^T D_s\, \mu^\circ(ds)\Big|\cF_t\Big],\quad 0\leq t\leq T
    \end{equation}
    and in particular $L_t\leq k_2 \big(\mu^\circ[t,T]\big)^{1-p}$ if $D\leq k_2$.
    In the case without intermediate consumption, $L$ is a submartingale.

    If there exists an optimal strategy $(\hpi,\hc)$, then $L,L_->0$.
  \end{enumerate}
\end{Lemma}

\begin{proof}
  Consider the cases where either $p>0$, or $p<0$ \emph{and} there is no intermediate consumption. Then $\pi\equiv0$, $c\equiv x_01_{\{T\}}$ is an admissible strategy and Proposition~\ref{pr:martOptPrincipleForL} shows that $L_t\tfrac{1}{p}x_0^p+\int_0^t U_s(0)\,\mu(ds)=L_t\tfrac{1}{p}x_0^p$ is a supermartingale, proving the  super/submartingale properties in~(i) and~(ii).

  Let $p$ be arbitrary and assume there is no intermediate consumption. Applying~\eqref{eq:OppProcIndep} with $\pi\equiv0$ and $c\equiv x_01_{\{T\}}$, we get
  $L_t \tfrac{1}{p} x_0^p\geq E[U_T(c_T)|\cF_t] = E[D_T|\cF_t]\tfrac{1}{p} x_0^p$.
  Hence $L_t\geq E[D_T|\cF_t]$ if $p>0$ and $L_t\leq E[D_T|\cF_t]$ if $p<0$, which corresponds to~\eqref{eq:BoundForL1} and
  \eqref{eq:BoundForL2} for this case.

  If there is intermediate consumption (and $p$ is arbitrary), we consume at a constant rate after the fixed time $t$. That is,
   we use~\eqref{eq:OppProcIndep} with $\pi\equiv 0$ and $c=x_0(T-t+1)^{-1} 1_{[t,T]}$ to obtain
  $L_t\tfrac{1}{p}x_0^p\geq E\big[\int_t^T U_s(c_s)\,\mu^\circ(ds)\big|\cF_t\big]
       =\tfrac{1}{p}x_0^p (1+T-t)^{-p} E\big[\int_t^T D_s \,\mu^\circ(ds)\big|\cF_t\big]$.
  This ends the proof of~\eqref{eq:BoundForL1} and
  \eqref{eq:BoundForL2}.

  In the case $p<0$,~\eqref{eq:BoundForL2} shows that $L$ is dominated by a martingale, hence $L$ is of class (D) and in particular a special semimartingale.

  It remains to prove the positivity. If $p>0$,~\eqref{eq:BoundForL1} shows $L>0$ and then $L_->0$ follows by the minimum principle for positive supermartingales. For $p<0$, let $\hX=X(\hpi,\hc)$ be the optimal wealth process. Clearly $L>0$ follows from~\eqref{eq:OppProcIndep} with $(\hpi,\hc)$. From Proposition~\ref{pr:martOptPrincipleForL} we have that $\tfrac{1}{p}\hX^p L +\int U_s(\hc_s)\,\mu(ds)$ is a negative martingale, hence $\hX^p L$ is a positive supermartingale.
  Therefore $P[\inf_{0\leq t\leq T} \hX^p_t L_t >0]=1$ and it remains to note that the paths of $\hX^p$ are $P$-a.s.~bounded because $\hX,\hX_->0$.
\end{proof}

The following concerns the submartingale property in Lemma~\ref{le:BoundsForL}(ii).

\begin{Example}\label{ex:LcanBeSupermart}
  Consider the case \emph{with} intermediate consumption and
  assume that $D\equiv 1$ and $S\equiv 1$. Then
  $(\hpi,\hc)\equiv (0,x_0/(1+T))$ is an optimal strategy and $L_t=(1+T-t)^{1-p}$ is a decreasing function. In particular, $L$ is not a submartingale.
\end{Example}

\begin{Remark}\label{rk:constraintsDef}
  We can also consider the utility maximization problem under \emph{constraints} in the following sense. Suppose that for each
  $(\omega,t)\in\Omega\times [0,T]$ we are given a set $\sC_t(\omega)\subseteq \R^d$. We assume that each of these sets contains the origin.
  A strategy $(\pi,c)\in\cA$ is called \emph{$\sC$-admissible} if $\pi_t(\omega)\in\sC_t(\omega)$ for all $(\omega,t)$, and the set
  of all these strategies is denoted by $\cA^\sC$. %
  We remark that all results (and their proofs) in this section remain valid if $\cA$ is replaced by $\cA^\sC$ throughout.
  This generalization is not true for the subsequent section and existence of an optimal strategy is not guaranteed for general $\sC$.
\end{Remark}

\section{Relation to the Dual Problem}\label{se:OppProcAndDuality}
We discuss how the problem dual to utility maximization relates to the opportunity process $L$. \textbf{We assume~\eqref{eq:BoundsR} and~\eqref{eq:ELMMexists} in the entire Section~\ref{se:OppProcAndDuality}}, hence Proposition~\ref{pr:ExistenceKZ} applies.
The dual problem will be defined on a domain $\sY$ introduced below. Since its definition is slightly cumbersome,
we point out that to follow the results in the body of this paper, only two facts about $\sY$ are needed. First, the density
process of each martingale measure $Q\in\sM^{S}$, scaled by a certain constant $y_0$, is contained in $\sY$. Second, each element of $\sY$ is a positive supermartingale.

Following~\cite{KaratzasZitkovic.03}, the \emph{dual problem} is given by
\begin{equation}\label{eq:dualProblem}
  \inf_{Y\in\sY(y_0)} E\Big[\int_0^T U_t^*(Y_t)\,\mu^\circ(dt)\Big],
\end{equation}
where $y_0:=u'(x_0)=L_0 x_0^{p-1}$ and $U^*_t$ is the convex conjugate of $x\mapsto U_t(x)$,
\begin{equation}\label{eq:convexConjugate}
  U_t^*(y):=\sup_{x>0} \big\{U_t(x)-xy\big\}=-\tfrac{1}{q}y^{q}D_t^{\beta}.
\end{equation}
We have denoted by
\begin{equation}\label{eq:betaAndq}
  \beta:=\frac{1}{1-p}>0,\quad  q:=\frac{p}{p-1}\in(-\infty,0)\cup (0,1)
\end{equation}
the relative risk tolerance and the exponent conjugate to $p$, respectively. These constants will be used very often in the sequel and it is useful to note $\sgn(p)=-\sgn(q)$.
It remains to define the domain $\sY=\sY(y_0)$. Let
\[
  \sX=\{H\sint S:\, H\in L(S),\; H\sint S \mbox{ is bounded below}\}
\]
be the set of gains processes from trading. The set of ``supermartingale densities'' is defined by
\[
  \sY^*=\{Y\geq 0 \mbox{ c\`adl\`ag}:\, Y_0\leq y_0,\;\; YG \mbox{ is a supermartingale for all }G\in\sX\};
\]
its subset corresponding to probability measures equivalent to $P$ on $\cF_T$ is
\[
  \sY^{\sM}=\{Y\in \sY^*:\, Y>0\mbox{ is a martingale and }Y_0=y_0\}.
\]
We place ourselves in the setting of~\cite{KaratzasZitkovic.03} by considering the same dual domain $\sY^{\sD}\subseteq\sY^*$. It consists of the density processes of (the regular parts of) the finitely additive measures
contained in the $\sigma((L^\infty)^*,L^\infty)$-closure of the set $\{Y_T:\,Y\in\sY^{\sM}\}\subseteq L^1\subseteq (L^{\infty})^*$. More precisely, we multiply each density with the constant $y_0$. We refer to~\cite{KaratzasZitkovic.03} for details as the precise construction of $\sY^{\sD}$ is not important here, it is relevant for us only that $\sY^{\sM}\subseteq\sY^{\sD}\subseteq\sY^*$. In particular, $y_0 \sM^S \subseteq \sY^{\sD}$ if we identify measures and their density processes.
For notational reasons, we make the dual domain slightly smaller and let
\[
 \sY:=\{Y\in\sY^{\sD}:\,Y>0\}.
\]
By~\cite[Theorem~3.10]{KaratzasZitkovic.03} there exists a unique $\hY=\hY(y_0)\in\sY$ such that the infimum in~\eqref{eq:dualProblem} is attained, and it is related to the optimal consumption $\hc$ via the marginal utility by
\begin{equation}\label{eq:DualOptAndConsumption}
  \hY_t=\partial_x\,U_t(x)|_{x=\hc_t}=D_t \hc_t^{p-1}
\end{equation}
on the support of $\mu^\circ$. In the case without intermediate consumption, an existence result was previously obtained in~\cite{KramkovSchachermayer.99}.

\begin{Remark}
  All the results stated below remain true if
  we replace $\sY$ by $\{Y\in\sY^*:\,Y>0\}$; i.e., it is not important for our purposes whether we use the dual domain
  of~\cite{KaratzasZitkovic.03} or the one of~\cite{KramkovSchachermayer.99}.
  This is easily verified using the fact that $\sY^{\sD}$ contains all maximal elements of
  $\sY^*$ (see \cite[Theorem~2.10]{KaratzasZitkovic.03}). Here $Y\in \sY^*$ is called maximal if $Y=Y'B$, for some $Y'\in\sY^*$ and some c\`adl\`ag nonincreasing process $B\in[0,1]$, implies $B\equiv 1$.
\end{Remark}

\begin{Prop}\label{pr:dualMinimizer}
  Let $(\hc,\hpi)\in\cA$ be an optimal strategy and $\hX=X(\hpi,\hc)$.
  The solution to the dual problem is given by
  \[
    \hY=L \hX^{p-1}.
  \]
\end{Prop}

\begin{proof}
  As $L_T=D_T$ and $\hc_T=\hX_T$,~\eqref{eq:DualOptAndConsumption} already yields $\hY_T=L_T \hX^{p-1}_T$. Moreover,
   by Lemma~\ref{le:MartingaleOfDuality} in the Appendix, $\hY$ has the property that
  \[
    Z_t:=\hY_t\hX_t+\int_0^t\hY_s\hc_s\,\mu(ds)=\hY_t\hX_t+p\int_0^t U_s(\hc_s)\,\mu(ds)
  \]
   is a martingale.
   By Proposition~\ref{pr:martOptPrincipleForL},
   $
    \widetilde{Z}_t:=L_t \hX_t^p + p\int_0^t U_s(\hat{c}_s)\,\mu(ds)
   $
   is also a martingale. The terminal values of these martingales coincide, hence $\widetilde{Z}=Z$. We deduce $\hY=L \hX^{p-1}$ as $\hX>0$.
\end{proof}

The formula $\hY=L \hX^{p-1}$ could be used to \emph{define} the opportunity process $L$. This is the approach taken in Muhle-Karbe~\cite{MuhleKarbe.09} (see also ~\cite{KallsenMuhleKarbe.08}), where utility from terminal wealth is considered and the opportunity process is used as a tool to verify the optimality of an explicit candidate solution.
From a systematic point of view, our approach via the value process has the advantage that it immediately yields the properties in Lemma~\ref{le:BoundsForL}
and certain monotonicity results (see Section~\ref{se:ApplicationsOfFeedbackFormula}).

\subsection{The Dual Opportunity Process}

Since the function $U^*$ in the dual problem~\eqref{eq:dualProblem} is again homogeneous, we expect a similar structure as in the primal problem. This is formalized by the dual opportunity process $L^*$. Not only is it natural to introduce this object, it also turns out that in certain situations $L^*$ is a more convenient tool than $L$  (e.g.,~\cite{Nutz.09d}). %
We define for $Y\in\sY$ and $t\in [0,T]$ the set %
\[
  \sY(Y,t):=\big\{\widetilde{Y}\in\sY:\, \widetilde{Y}=Y\mbox{ on } [0,t]\big\}
\]
and we recall the constants~\eqref{eq:betaAndq} and the standing assumptions~\eqref{eq:BoundsR} and~\eqref{eq:ELMMexists}.

\begin{Prop}\label{pr:DualOppProc}
  There exists a unique c\`adl\`ag process $L^*$, called \emph{dual opportunity process}, such that
  for all $Y\in\sY$ and $t\in [0,T]$,
  \[
    -\tfrac{1}{q}Y_t^q L^*_t
    = \mathop{\essinf}_{\widetilde{Y}\in\sY(Y,t)} E\Big[\int_t^T U_s^*(\widetilde{Y}_s)\,\mu^\circ(ds)\Big|\cF_t\Big].
  \]
  An alternative description is
  \[
    L^*_t =
  \begin{cases}
    \mathop{\esssup}_{Y\in\sY}& \hspace{-.8em} E\Big[\int_t^T D_s^\beta (Y_s/Y_t)^q\,\mu^\circ(ds)\Big|\cF_t\Big]
    \;\text{ if }q\in(0,1),\phantom{\bigg|}\\
    \mathop{\essinf}_{Y\in\sY}& \hspace{-.8em} E\Big[\int_t^T D_s^\beta (Y_s/Y_t)^q\,\mu^\circ(ds)\Big|\cF_t\Big]
    \;\text{ if }q<0
  \end{cases}
  \]
  and the extrema are attained at $Y=\hY$.
\end{Prop}

\begin{proof}
  The fork convexity of $\sY$ \cite[Theorem~2.10]{KaratzasZitkovic.03} shows that if $Y,\check{Y} \in\sY$ and $\tY\in\sY(\check{Y},t)$, then $Y1_{[0,t)}+(Y_t/\check{Y}_t)\tY1_{[t,T]}$ is in $\sY(Y,t)$. It also implies that
  if $A\in\cF_t$ and $Y^1,Y^2\in\sY(Y,t)$, then $Y^1 1_A+Y^2 1_{A^c}\in\sY(Y,t)$.
  The proof of the first claim is now analogous to that of Proposition~\ref{pr:OppProcIndep}.
  The second part follows by using that $L^*$ does not depend on $Y$.
\end{proof}

The process $L^*$ is related to $L$ by a simple power transformation.

\begin{Prop}\label{pr:LBetaDualFormula}
  Let $\beta=\frac{1}{1-p}$. Then $L^*=L^\beta$.
\end{Prop}

\begin{proof}

  The martingale property of $Z_t:=\hX_t\hY_t+ \int_0^t \hc_s\hY_s\,\mu(ds)$ from Lemma~\ref{le:MartingaleOfDuality}
  implies that $\hX_t\hY_t=E[Z_T|\cF_t] - \int_0^t \hc_s\hY_s\,\mu(ds)
  = E\big[\int_t^T \hc_s\hY_s\,\mu^\circ(ds)\big|\cF_t\big] = E\big[\int_t^T D_s^\beta \hY^q_s\,\mu^\circ(ds)\big|\cF_t\big]$,
  where the last equality is obtained by expressing $\hc$ via~\eqref{eq:DualOptAndConsumption}. The right hand side
  equals $\hY^{q}_t L^*_t$ by Proposition~\ref{pr:DualOppProc}; so we have shown $\hX\hY=\hY^q L^*$.
  On the other hand, $(L\hX^{p-1})^q=\hY^q$ by Proposition~\ref{pr:dualMinimizer} and this can be written as
  $\hX\hY=\hY^q L^\beta$. We deduce $L^*=L^\beta$ as $\hY>0$.
\end{proof}

\subsection{Reverse H\"older Inequality and Boundedness of $L$}

In this section we study uniform bounds for $L$ in terms of inequalities of reverse H\"older type. %
This will yield a corresponding result for the optimal consumption in Section~\ref{se:ApplicationsOfFeedbackFormula} as well as a sufficient condition for~\eqref{eq:PrimalProblemFinite}. Moreover,
uniform bounds for $L$ are linked to the existence of bounded solutions for a certain class of backward stochastic differential equations, as explained in the companion paper~\cite{Nutz.09b}. Since bounded solutions are of particular interest in the theory of those equations, the detailed treatment below is also motivated by this link. %

Let $q=\frac{p}{p-1}$ be the exponent conjugate to $p$. Given a general positive process $Y$, we consider the following inequality:
\begin{equation}\label{eq:ReverseHolder}
  \begin{cases}
    \displaystyle{\int_\tau^T} E\big[(Y_s/Y_{\tau})^q\big|\cF_\tau\big]\,\mu^\circ(ds)\leq C_q \q& \text{if }q<0, \phantom{\Bigg|} \\
    \displaystyle{\int_\tau^T} E\big[(Y_s/Y_{\tau})^q\big|\cF_\tau\big]\,\mu^\circ(ds)\geq C_q \q& \text{if }q\in(0,1),
  \end{cases}\tag{$\mathrm{R}_q(P)$}
\end{equation}
for all stopping times $0\leq\tau\leq T$ and some constant $C_q>0$ independent of~$\tau$.  It is useful to recall
that $q<0$ corresponds to $p\in (0,1)$ and vice versa.

Without consumption,~\ref{eq:ReverseHolder} reduces to $E[(Y_T/Y_{\tau})^q|\cF_\tau]\leq C_q$ (resp. the converse inequality).
Inequalities of this type are well known. See, e.g., Dol\'{e}ans-Dade and Meyer~\cite{DoleansDadeMeyer.79}
for an introduction or Delbaen et al.~\cite{DelbaenEtAl.97} and the references therein for some connections to finance. In most applications, the considered exponent $q$ is greater than one;
\ref{eq:ReverseHolder} then takes the form as for $q<0$.
We recall once more the standing assumptions~\eqref{eq:BoundsR} and~\eqref{eq:ELMMexists}.

\begin{Prop}\label{pr:RevHolderIneqAndL}
  The following are equivalent:\nopagebreak
  \begin{enumerate}[topsep=3pt, partopsep=0pt, itemsep=1pt,parsep=2pt]
    \item The process $L$ is uniformly bounded away from zero and infinity.
    \item Inequality~\ref{eq:ReverseHolder} holds for the dual minimizer $\hY\in\sY$.
    \item Inequality~\ref{eq:ReverseHolder} holds for \emph{some} $Y\in\sY$.
  \end{enumerate}
\end{Prop}

\begin{proof}
  Under the standing assumption~\eqref{eq:BoundsR}, a one-sided bound for $L$ always holds by
  Lemma~\ref{le:BoundsForL}, namely $L\geq k_1$ if $p\in (0,1)$ and $L\leq const.$ if $p<0$.

  (i) is equivalent to (ii): We use~\eqref{eq:BoundsR} and then Propositions~\ref{pr:DualOppProc} and~\ref{pr:LBetaDualFormula} to obtain that
  $\int_\tau^T E\big[(\hY_s/\hY_\tau)^q\big|\cF_\tau\big]\,\mu^\circ(ds)
  = E\big[\int_\tau^T(\hY_s/\hY_\tau)^q \,\mu^\circ(ds) \big|\cF_\tau\big]
  \leq k_1^{-\beta}E\big[\int_\tau^T D_s^\beta (\hY_s/\hY_\tau)^q \,\mu^\circ(ds) \big|\cF_\tau\big]
  =k_1^{-\beta}L_\tau^*=k_1^{-\beta}L_\tau^\beta$.
    Thus when $p\in(0,1)$ and hence $q<0$,~\ref{eq:ReverseHolder} for $\hY$ is equivalent to an upper bound for $L$. For $p<0$, we replace $k_1$ by $k_2$.

  (iii) implies (i): Assume $p\in(0,1)$. Using Propositions~\ref{pr:LBetaDualFormula} and~\ref{pr:DualOppProc} and~\eqref{eq:convexConjugate},
  $-\tfrac{1}{q}Y_t^q L^{\beta}_t
   \leq E\big[\int_t^T U_s^*(Y_s)\,\mu^\circ(ds)\big|\cF_t\big] \leq -\tfrac{1}{q} k_2^\beta \int_t^T E[Y^q_s|\cF_t]\,\mu^\circ(ds)$.
  Hence $L\leq k_2 C_q^{-\beta}$. If $p<0$, we obtain $L\geq k_1 C_q^{-\beta}$ in the same way.
\end{proof}

If the equivalent conditions of Proposition~\ref{pr:RevHolderIneqAndL} are satisfied, we say that ``\ref{eq:ReverseHolder} holds'' for the given financial market model.
Although quite frequent in the literature, this condition is rather restrictive in the sense that it often fails in explicit models that have stochastic dynamics.
For instance, in the affine models of~\cite{KallsenMuhleKarbe.08}, $L$ is an exponentially affine function of a typically unbounded factor process, in which case Proposition~\ref{pr:RevHolderIneqAndL} implies that~\ref{eq:ReverseHolder} fails. Similarly, $L$ is an exponentially quadratic function of an Ornstein-Uhlenbeck process in the model of Kim and Omberg~\cite{KimOmberg.96}. On the other hand, exponential L\'evy models have constant dynamics and here $L$ turns out to be
simply a smooth deterministic function (see Example~\ref{ex:levy}).

In a given model, it may be hard to check whether~\ref{eq:ReverseHolder} holds. Recalling $y_0\sM^S\subseteq \sY$, an obvious approach in view of Proposition~\ref{pr:RevHolderIneqAndL}(iii) is to choose for $Y/y_0$ the density process of some specific martingale measure. We illustrate this with an essentially classical example.

\begin{Example}\label{ex:bddMeanVarianceTradeoff}
  Assume that $R$ is a special semimartingale with decomposition
  \begin{equation}\label{eq:StructureCondJumps}
    R=\alpha\sint \br{R^c}+M^R,
  \end{equation}
  where $R^c$ denotes the continuous local martingale part of $R$, $\alpha\in L^2_{loc}(R^c)$, and $M^R$ is the local martingale part of $R$.
  Suppose that the process
  \[
    \chi_t:=\int_0^t \alpha_s^\top \,d\br{R^c}_s \,\alpha_s \, ,\q t\in [0,T]
  \]
  is \emph{uniformly bounded}. Then $Z:=\cE(-\alpha \sint R^c)$ is a martingale by Novikov's condition and the measure $Q\approx P$ with density $dQ/dP=Z_T$ is a local martingale measure for $S$ as $Z\cE(R)=\cE(-\alpha\sint R^c + M^R)$ by Yor's formula; hence $y_0 Z\in\sY$. Fix $q$. Using
  $
    Z^q=\cE(-q \alpha \sint R^c)\exp\big(\tfrac{1}{2}q(q-1) \chi\big),
  $
  and that $\cE(-q \alpha \sint R^c)$ is a martingale by Novikov's condition, one readily checks that $Z$ satisfies inequality~\ref{eq:ReverseHolder}.

  If $R$ is continuous,~\eqref{eq:StructureCondJumps} is the structure condition of Schweizer~\cite{Schweizer.95b} and under~\eqref{eq:ELMMexists} $R$ is necessarily of this form. Then $\chi$ is called mean-variance tradeoff process and $Q$ is the ``minimal'' martingale measure.
  In It\^o process models, $\chi$ takes the form $\chi_t=\int_0^t \theta^\top_s\theta_s\,ds$, where $\theta$ is the market price of risk process.
  Thus $\chi$ will be bounded whenever $\theta$ is.
\end{Example}

\begin{Remark}\label{rk:suffCondFiniteness}
  The example also gives a sufficient condition for~\eqref{eq:PrimalProblemFinite}. This is of interest only for $p\in (0,1)$ and we remark that for the case of It\^o process models with bounded $\theta$, the condition corresponds to Karatzas and Shreve~\cite[Remark~6.3.9]{KaratzasShreve.98}.

  Indeed, if there exists $Y\in\sY$ satisfying~\ref{eq:ReverseHolder}, then with~\eqref{eq:convexConjugate} and~\eqref{eq:BoundsR} it follows that the
  the value of the dual problem~\eqref{eq:dualProblem} is finite, and this suffices for~\eqref{eq:PrimalProblemFinite}, as in Kramkov and Schachermayer~\cite{KramkovSchachermayer.03}.
\end{Remark}

The rest of the section studies the dependence of $\mathrm{R}_q(P)$ on $q$.

\begin{Remark}\label{rk:unifRevHolder}
  Assume that $Y$ satisfies $\mathrm{R}_q(P)$ with a constant $C_q$. If $q_1$ is such that
  $q<q_1<0$ or $0<q<q_1<1$, then $\mathrm{R}_{q_1}(P)$ is satisfied with
  \[
   C_{q_1}=\big(\mu^\circ[0,T]\big)^{1-q_1/q} (C_q)^{q_1/q}.
  \]
  Similarly, if $q<0<q_1<1$, we can take $C_{q_1}=(C_q)^{q_1/q}$.
  This follows from Jensen's inequality.
\end{Remark}

There is also a partial converse.

\begin{Lemma}\label{le:revHolderConverse}
  Let $0<q<q_1<1$ and let $Y>0$ be a supermartingale. If $Y$ satisfies $\mathrm{R}_{q_1}(P)$, it also satisfies
  \ref{eq:ReverseHolder}.

  In particular, the following dichotomy holds: $Y$ satisfies either all or none of the inequalities~$\big\{$\ref{eq:ReverseHolder}, $q\in (0,1)\big\}$.
\end{Lemma}

\begin{proof}
  From Lemma~\ref{le:monotoneFunctionRevHolder} stated below  we have
  $\int_t^T E\big[(Y_s/Y_t)^{q} \big|\cF_t\big]\,\mu^\circ(ds)
     \geq \int_t^T \big(E\big[(Y_s/Y_t)^{q_1}\big|\cF_t\big]\big)^{\frac{1-q}{1-q_1}}\,\mu^\circ(ds)$.
  Noting that $\frac{1-q}{1-q_1}>1$, we apply Jensen's inequality to the right-hand side
   and then use $\mathrm{R}_{q_1}(P)$ to deduce the claim with $C_q:=\big(\mu^\circ[t,T]\big)^{\frac{q-q_1}{1-q_1}}\,(C_{q_1})^{\frac{1-q}{1-q_1}}$.
  The dichotomy follows by the previous remark.
\end{proof}

For future reference, we state separately the main step of the above proof.

\begin{Lemma}\label{le:monotoneFunctionRevHolder}
  Let $Y>0$ be a supermartingale. For fixed $0\leq t\leq s\leq T$, %
  \[
    \phi: \, (0,1)\to \R_+,\quad q\mapsto \phi(q):=\Big(E\big[(Y_s/Y_t)^q\big|\cF_t\big]\Big)^{\frac{1}{1-q}}
  \]
  is a monotone decreasing function $P$-a.s. If in addition $Y$ is a martingale, then
  $\lim_{q\to 1-}\phi(q)=\exp\big(-E\big[(Y_s/Y_t)\log(Y_s/Y_t)\big|\cF_t\big]\big)$ $P$-a.s.,
  where the conditional expectation has values in $\R\cup\{+\infty\}$.
\end{Lemma}

\begin{proof}
   Suppose first that $Y$ is a martingale; by scaling we may assume $E[Y_.]=1$.  We define a probability $Q\approx P$ on $\cF_s$ by $dQ/dP:=Y_s$. With $r:=(1-q)\in (0,1)$ and Bayes' formula,
   \[
     \phi(q)=\Big(Y_t^{1-q} E^Q\big[Y_s^{q-1}\big|\cF_t\big]\Big)^{\frac{1}{1-q}}
     =Y_t \Big(E^Q\big[(1/Y_s)^{r}\big|\cF_t\big]\Big)^{\frac{1}{r}}.
   \]
   This is increasing in $r$ by Jensen's inequality, hence decreasing in $q$.

   Now let $Y$ be a supermartingale. We can decompose it as $Y_u=B_uM_u$, $u\in[0,s]$, where $M$ is a martingale and $B_s=1$. That is,
   $M_t=E[Y_s|\cF_t]$ and $B_t=Y_t/E[Y_s|\cF_t]\geq 1$, by the supermartingale property. Hence $B_t^{q/(q-1)}$ is decreasing in $q\in(0,1)$. Together with the first part, it follows that
   $\phi(q)=B_t^{q/(q-1)}\big(E\big[(M_s/M_t)^q\big|\cF_t\big]\big)^{\frac{1}{1-q}}$ is decreasing.

   Assume again that $Y$ is a martingale. The limit $\lim_{q\to 1-} \log\big(\phi(q)\big)$ can be calculated as
   \[
     \lim_{q\to 1-} \frac{\log\big(E\big[(Y_s/Y_t)^q\big|\cF_t\big]\big)}{1-q}
     = \lim_{q\to 1-} -\frac{E\big[(Y_s/Y_t)^q \log(Y_s/Y_t)\big|\cF_t\big]}{E\big[(Y_s/Y_t)^q\big|\cF_t\big]}\quad P\mbox{-a.s.}
   \]
   using l'H\^opital's rule and $E[(Y_s/Y_t)|\cF_t]=1$. The result follows using monotone and bounded convergence in the
   numerator and dominated convergence in the denominator.
\end{proof}

\begin{Remark}
  The limiting case $q=1$ corresponds to the entropic inequality
  $\mathrm{R}_{L\log L}(P)$ which reads $\int_\tau^T E\big[(Y_s/Y_{\tau})\log(Y_s/Y_{\tau})\big|\cF_\tau\big]\,\mu^\circ(ds)\leq C_1$. Lemma~\ref{le:monotoneFunctionRevHolder} shows that for a martingale $Y>0$, $\mathrm{R}_{q_1}(P)$ with $q_1\in (0,1)$ is weaker than
  $\mathrm{R}_{L\log L}(P)$, which, in turn, is obviously weaker than $\mathrm{R}_{q_0}(P)$ with $q_0>1$.

  A much deeper argument~\cite[Proposition~5]{DoleansDadeMeyer.79} shows that if $Y$ is a martingale satisfying
  the ``condition (S)'' that $k^{-1} Y_-\leq Y\leq kY_-$ for some $k>0$,  then $Y$ satisfies
  $\mathrm{R}_{q_0}(P)$ for some $q_0>1$ if and only if it satisfies $\mathrm{R}_{q}(P)$ for some  $q<0$,
  and then by Remark~\ref{rk:unifRevHolder} also $\mathrm{R}_{q_1}(P)$ for all $q_1\in (0,1)$.
\end{Remark}

Coming back to the utility maximization problem, we obtain the following dichotomy
from Lemma~\ref{le:revHolderConverse} and the implication (iii)~$\Rightarrow$~(ii) in Proposition~\ref{pr:RevHolderIneqAndL}.

\begin{Cor}\label{co:revHolderDichotomy}
  For the given market model, $\mathrm{R}_{q}(P)$ holds  either for all or no values of $q\in (0,1)$.
\end{Cor}

We believe that this equivalence of reverse H\"older inequalities is surprising and also of independent probabilistic interest. %

\section{Applications}\label{se:ApplicationsOfFeedbackFormula}

In this section we consider only the case \emph{with} intermediate consumption. \textbf{We assume~\eqref{eq:BoundsR} and~\eqref{eq:ELMMexists}}. However, we remark that all results except for Proposition~\ref{pr:kappaBounds} and Remark~\ref{rk:oppositeBoundsKappa} hold true as soon as there exists an optimal strategy $(\hpi,\hc)\in\cA$.

We first show the announced feedback formula for the optimal propensity to consume $\hkappa$, which will then allow us to translate the results of the previous sections into economic statements. The following theorem can be seen as a generalization
of~\cite[Proposition~8]{StoikovZariphopoulou.05}, which considers a Markovian model with It\^o coefficients driven by a correlated factor.

\begin{Thm}\label{th:consumptionFeedback}
  With $\beta=\tfrac{1}{1-p}$ we have
  \begin{equation}\label{eq:consumptionFeedback}
    \hc_t=\Big(\frac{D_t}{L_t}\Big)^\beta\hX_t \q\q\mbox{and hence}\q\q \hkappa_t=\Big(\frac{D_t}{L_t}\Big)^\beta.
  \end{equation}
\end{Thm}
\begin{proof}
  This follows from Proposition~\ref{pr:dualMinimizer} via~\eqref{eq:DualOptAndConsumption} and~\eqref{eq:DefPropToConsume}.
\end{proof}

\begin{Remark}\label{rk:FeedbackUnderConstraints}
  In~\cite[Theorem~3.2, Remark~3.6]{Nutz.09b} we generalize the formula for $\hkappa$ to the utility maximization problem under constraints as described in Remark~\ref{rk:constraintsDef}, under the sole assumption that an optimal constrained strategy exists. The proof relies on different techniques and is beyond the scope of this paper; we merely mention that $\hkappa$ is unique also in that setting.

  The special case where the constraints set $\sC\subseteq \R^d$ is linear can be deduced from Theorem~\ref{th:consumptionFeedback} by redefining the price process $S$. For instance, set $S^1\equiv1$ for $\sC=\{(x^1,\dots,x^d)\in\R^d:\, x^1=0\}$.
\end{Remark}

In the remainder of the section we discuss how certain changes in the model and
the discounting process $D$ affect the optimal propensity to consume. This is based on~\eqref{eq:consumptionFeedback}
and the relation
\begin{equation}\label{eq:restatedOppProcIndep}
  \tfrac{1}{p}x_0^p L_t=\mathop{\esssup}_{c\in\cA(0,x_01_{\{T\}},t)} E\Big[\int_t^T D_s\tfrac{1}{p}c_s^p\,\mu^\circ(ds)\Big|\cF_t\Big],
\end{equation}
which is immediate from Proposition~\ref{pr:OppProcIndep}. In the present non-Markovian setting the parametrization by the
propensity to consume is crucial as one cannot make statements for ``fixed wealth''. There is no immediate way to infer results about $\hc$, except of course for the initial value $\hc_0=\hkappa_0 x_0$.

\subsection{Variation of the Investment Opportunities}

It is classical in economics to compare two ``identical'' agents with utility function $U$, where only one has access to a stock market. The opportunity to invest in risky assets gives rise to two contradictory effects. The presence of risk incites the agent to save cash for the uncertain future; this is the \emph{precautionary savings effect} and its strength is related to the \emph{absolute prudence} $\mathscr{P}(U)=-U'''/U''$. On the other hand,
the agent may prefer to invest rather than to consume immediately. This \emph{substitution effect} is related to the \emph{absolute risk aversion} $\mathscr{A}(U)=-U''/U'$.

Classical economic theory (e.g., Gollier~\cite[Proposition~74]{Gollier.01}) states that in a one period model, the presence of a complete financial market makes the optimal consumption at time $t=0$ smaller if $\mathscr{P}(U)\geq 2 \mathscr{A}(U)$ holds everywhere on $(0,\infty)$, and larger if the converse inequality holds.
For power utility, the former condition holds if $p<0$ and the latter holds if $p\in (0,1)$.
We go a step further in the comparison by considering two different sets of constraints, instead of
giving no access to the stock market at all (which is the constraint $\{0\}$).

Let $\sC$ and $\sC'$ be set-valued mappings of constraints as in Remark~\ref{rk:constraintsDef},
and let $\sC'\subseteq \sC$ in the sense that $\sC'_t(\omega)\subseteq \sC_t(\omega)$ for all $(t,\omega)$.
\emph{Assume} that there exist corresponding optimal constrained strategies.

\begin{Prop}\label{pr:kappaMonotoneConstraints}
  Let $\hkappa$ and $\hkappa'$ be the optimal propensities to consume for the constraints $\sC$ and $\sC'$, respectively.
  Then $\sC'\subseteq \sC$ implies $\hkappa\leq \hkappa'$ if $p>0$ and $\hkappa\geq \hkappa'$ if $p<0$.
  In particular, $\hc_0\leq \hc_0'$ if $p>0$ and $\hc_0\geq \hc_0'$ if $p<0$.
\end{Prop}

\begin{proof}
  Let $L$ and $L'$ be the corresponding opportunity processes;
  we make use of Remarks~\ref{rk:constraintsDef} and~\ref{rk:FeedbackUnderConstraints}.
  Consider relation~\eqref{eq:restatedOppProcIndep} with $\cA^{\sC}$ instead of $\cA$ and the analogue for $L'$ with
  $\cA^{\sC'}$. We see that $\cA^{\sC'}\subseteq\cA^{\sC}$ implies $\tfrac{1}{p}L'\leq \tfrac{1}{p}L$, as the supremum is taken over a larger set in the case of $\sC$. By~\eqref{eq:consumptionFeedback}, $\hkappa$ is a decreasing function of $L$.
\end{proof}

\begin{Prop}\label{pr:kappaBounds}
 The optimal propensity to consume satisfies
  \[
   \hkappa_t \leq \frac{(k_2/k_1)^\beta}{1+T-t} \;\mbox{ if }p\in (0,1) \quad\mbox{and}\quad \hkappa_t \geq \frac{(k_2/k_1)^\beta}{1+T-t} \;\mbox{ if }p<0.
 \]
 In particular, we have a model-independent deterministic threshold independent of $p$ in the standard case $D\equiv 1$,
 \[
   \hkappa_t \leq \frac{1}{1+T-t} \;\mbox{ if }p\in (0,1) \quad\mbox{and}\quad \hkappa_t \geq \frac{1}{1+T-t} \;\mbox{ if }p<0.
 \]
\end{Prop}

\begin{proof}
  This follows from Lemma~\ref{le:BoundsForL} and~\eqref{eq:consumptionFeedback}. The second part
  can also be seen as special case of Proposition~\ref{pr:kappaMonotoneConstraints} with constraint set $\sC'=\{0\}$
  since then $\hkappa'=(1+T-t)^{-1}$ as in Example~\ref{ex:LcanBeSupermart}.
\end{proof}

The threshold $(1+T-t)^{-1}$ coincides with the optimal propensity to consume for the
$\log$-utility function (cf.~\cite{GollKallsen.03}), which formally corresponds to $p=0$. This suggests that the threshold is attained by $\hkappa(p)$ in the limit $p\to0$, a result proved in~\cite{Nutz.09d}.

\begin{Remark}\label{rk:oppositeBoundsKappa}
  Uniform bounds for $\hkappa$ \emph{opposite} to the ones in Proposition~\ref{pr:kappaBounds} exist if and only if~\ref{eq:ReverseHolder} holds for the given financial market model. Quantitatively, if $C_q>0$ is the constant for~\ref{eq:ReverseHolder}, then
   \[
   \hkappa_t \geq \Big(\frac{k_2}{k_1}\Big)^\beta \frac{1}{C_q} \;\mbox{ if }p\in (0,1) \quad\mbox{and}\quad \hkappa_t\leq \Big(\frac{k_1}{k_2}\Big)^\beta \frac{1}{C_q} \;\mbox{ if }p<0.
 \]
  This follows from~\eqref{eq:consumptionFeedback} and~\eqref{eq:BoundsR} by (the proof of) Proposition~\ref{pr:RevHolderIneqAndL}.
  In view of Corollary~\ref{co:revHolderDichotomy} we have the following dichotomy:
  $\hkappa=\hkappa(p)$ has a uniform upper bound either for all values of $p<0$, or for none of them.
\end{Remark}

\subsection{Variation of $D$}

We now study how $\hkappa$ is affected if we increase $D$ on some time interval $[t_1,t_2)$.
To this end, let $0\leq t_1<t_2\leq T$ be two fixed points in time and $\xi$ a bounded c\`adl\`ag adapted process
which is strictly positive and nonincreasing on $[t_1,t_2)$.
In addition to $U_t(x)=D_t\tfrac{1}{p}x^p$ we consider the utility random field
\[
  U'_t(x):=D'_t\tfrac{1}{p}x^p,\quad D':=\big(1+\xi 1_{[t_1,t_2)}\big) D.
\]

As an interpretation, recall the modeling of taxation by $D$ from Remark~\ref{rk:InterpretationsForD}. Then we want to find out how the agent reacts to a temporary change of the tax policy on $[t_1,t_2)$---in particular whether a reduction of the tax rate $\varrho:=D^{-1/p}-1$ stimulates consumption. %
For $p>0$, the next result shows this to be true during $[t_1,t_2)$, while the contrary holds before the policy change and there is no effect after $t_2$. An agent with $p<0$ reacts in the opposite way.
Remark~\ref{rk:InterpretationsForD} also suggests other interpretations of the same result.

\begin{Prop}\label{pr:sensitivityRinterval}
  Let $\hkappa$ and $\hkappa'$ be the optimal propensities to consume for $U$ and $U'$, respectively. Then
  \[
  \begin{cases}
    \hkappa'_t<\hkappa_t &\mbox{ if }\; t<t_1, \\
    \hkappa'_t>\hkappa_t &\mbox{ if }\; t\in [t_1,t_2),\\
    \hkappa'_t=\hkappa_t &\mbox{ if }\; t\geq t_2.
  \end{cases}
  \]
\end{Prop}

\begin{proof}
  Let $L$ and $L'$ be the opportunity processes for $U$ and $U'$. We consider~\eqref{eq:restatedOppProcIndep} and
  compare it with its analogue for $L'$, where $D$ is replaced by $D'$.
  As $\xi>0$, we then see that $L'_t>L_t$ for $t<t_1$; moreover, $L'_t=L_t$ for $t\geq t_2$. Since $\xi$ is nonincreasing,
  we also see that $L'_t<(1+\xi_t) L_t$ for $t\in [t_1,t_2)$.
  It remains to apply~\eqref{eq:consumptionFeedback}. For $t<t_1$, $\hkappa'=(D'_t/L'_t)^\beta=(D_t/L'_t)^\beta <(D_t/L_t)^\beta=\hkappa$.
  For $t\in [t_1,t_2)$ we have
  \[
   \hkappa'=(D'_t/L'_t)^\beta=\Big(\frac{(1+\xi_t) D_t}{L'_t}\Big)^\beta > \Big(\frac{(1+\xi_t) D_t}{(1+\xi_t) L_t}\Big)^\beta=\hkappa,
  \]
  while for $t\geq t_2$, $D'_t=D_t$ implies $\hkappa'_t=\hkappa_t$.
\end{proof}

\begin{Remark}
 (i)~~For $t_2=T$, the statement of Proposition~\ref{pr:sensitivityRinterval} remains true if the closed interval is chosen in the definition of $\widetilde{D}$.

 (ii)~~One can see~\cite[Proposition~12]{StoikovZariphopoulou.05} as a special case of Proposition~\ref{pr:sensitivityRinterval}.
  In our notation, the authors consider $D=1_{[0,T)}K_1+1_{\{T\}}K_2$ for two constants $K_1,K_2>0$
  and obtain monotonicity of the consumption with respect to the ratio $K_2/K_1$. This is proved in a Markovian setting by a comparison result for PDEs.
\end{Remark}

\section{On the Optimal Trading Strategy}\label{se:tradingStrategy}

In this section we indicate how the opportunity process $L$ describes the optimal trading strategy $\hpi$.
This issue is thoroughly treated in~\cite{Nutz.09b} and our aim here is only to complete the picture of how the opportunity process describes the power utility maximization problem. The following holds whenever an optimal strategy $(\hpi,\hkappa)$ exists  and in particular when the conditions of Proposition~\ref{pr:ExistenceKZ} are satisfied.

Our description for $\hpi$ is local, i.e., we fix $(\omega,t)\in \Omega\times [0,T]$ and characterize the vector
$\hpi_t(\omega)\in \R^d$. We shall see that this vector maximizes a certain concave function $g$, or more precisely, a function $y\mapsto g(\omega,t,y)$ on a subset $\sC^0_t(\omega)$ of $\R^d$. Therefore, $\hpi_t(\omega)$ can be seen as the optimal control for a deterministic control problem whose admissible controls are given by the set $\sC^0_t(\omega)$:
\begin{equation}\label{eq:optStrategyGen}
    \hpi_t(\omega) = \mathop{\argmax}_{y\in\sC^0_t(\omega)} g(\omega,t,y),\quad (\omega,t)\in \Omega\times [0,T].
\end{equation}
The set $\sC^0_t(\omega)$ is a local description for the budget constraint, i.e., the condition that the wealth process $X(\pi,\kappa)$ corresponding to some strategy $(\pi,\kappa)$ has to be positive.

To formally define $\sC^0_t(\omega)$, we first have to introduce the semimartingale characteristics of $R$ (cf.~\cite[Chapter~II]{JacodShiryaev.03} for background).
Let $h: \R^d\to\R^d$ be a cut-off function, i.e.,
$h$ is bounded and  $h(x)=x$ in a neighborhood of $x=0$.
Moreover, we fix a suitable increasing process $A$ and denote by $(b^{R},c^{R},F^{R};A)$ the differential characteristics of $R$ with respect to $A$ and $h$. In the special case where $R$ is a L\'evy process, one can choose $A_t=t$ and then
$(b^{R},c^{R},F^{R})$ is the familiar L\'evy triplet. In general, the triplet $(b^{R},c^{R},F^{R})$ is stochastic, but for fixed  $(\omega,t)$ the interpretation is similar as in the L\'evy case. In particular, $F^R_t(\omega)$ is a L\'evy measure on $\R^d$ and describes the jumps of $R$. With this notation we can define
\[
  \sC^0_t(\omega):=\Big\{y\in\R^d:\, F^R_t(\omega)\big\{x\in\R^d:\, y^\top x < -1\big\}=0\Big\}.
\]
This formula is related to the budget constraint because the stochastic exponential $X(\pi,\kappa)=x_0\cE\big(\pi\sint R - \kappa \sint \mu\big)$ is nonnegative if and only if the jumps
of its argument satisfy $\pi^\top \Delta R\geq -1$, and this condition is expressed by the above formula in a local way.
We refer to~\cite[\S\,2.4]{Nutz.09b} for a detailed discussion and references to related literature.

It remains to specify the objective function $g$ of the local optimization problem~\eqref{eq:optStrategyGen}. To this end, we consider the $\R^d\times \R$-valued semimartingale $(R,L)$. We denote a generic point in $\R^d\times \R$ by $(x,x')$ and by
$(b^{R,L},c^{R,L},F^{R,L};A)$ the joint differential characteristics of $(R,L)$ with respect to $A$ and the cut-off function
$(x,x')\mapsto (h(x),x')$; here the last coordinate does not require a truncation because $L$ is special (Lemma~\ref{le:BoundsForL}).
Suppressing $(\omega,t)$ in the notation, we can now define
  \begin{align*}%
     g(y)
       & := L_{-}y^\top \Big( b^R + \tfrac{c^{RL}}{L_{-}}+ \tfrac{(p-1)}{2} c^R y \Big) + \int_{\R^d\times\R} x' y^\top h(x)\,F^{R,L}(d(x,x')) \nonumber\\
       &\phantom{:=} + \int_{\R^d\times\R} (L_-+x') \big\{p^{-1}(1+y^\top x)^p - p^{-1} - y^\top h(x)\big\}\, F^{R,L}(d(x,x')).
   \end{align*}
One can check (cf.~\cite[Appendix~A]{Nutz.09b}) that $y\mapsto g(\omega,t,y)$ is a well defined concave function on $\sC^0_t(\omega)$ taking values in the extended real line. With the above notation, and under the assumption that an optimal strategy $(\hpi,\hkappa)$ exists, \cite[Theorem~3.2]{Nutz.09b} states that the local description~\eqref{eq:optStrategyGen} for $\hpi$ holds true $P\otimes A$-a.e.

We conclude by an illustration of this result in the L\'evy case (see~\cite{Nutz.09c} for details).

\begin{Example}\label{ex:levy}
  Let $R$ be a scalar L\'evy process with L\'evy triplet $(b^{R},c^{R},F^{R})$ such that $R$ is neither an increasing nor a decreasing process (then the no-arbitrage condition~\eqref{eq:ELMMexists} is satisfied). We assume that the price process $S=\cE(R)$ is strictly positive, or equivalently that $F^{R}(-\infty,-1]=0$. We consider the standard power utility function $U(x)=\frac{1}{p}x^p$ and focus on the case with intermediate consumption for simplicity of notation.
  For $p\in (0,1)$, it turns out (see~\cite[Corollary~3.7]{Nutz.09c}) that our standing assumption~\eqref{eq:PrimalProblemFinite} is satisfied if and only if $\int |x|^p 1_{\{|x|>1\}} \,F^R(dx)<\infty$, and for $p<0$ we have seen that the assumption is always satisfied.

  The L\'evy setting has the particular feature that the opportunity process and the function $g$ are deterministic. More precisely,
  $g(\omega,t,y)=L_t\fg(y)$ for the deterministic and time-independent function
  \begin{equation*}%
   \fg(y) :=y^\top b^R + \tfrac{(p-1)}{2} y^\top c^R y
            + \int_{\R^d} \big\{p^{-1}(1+y^\top x)^p - p^{-1} - y^\top h(x)\big\}\, F^R(dx).
  \end{equation*}
  Since $L$ is positive, maximizing $g$ is equivalent to maximizing $\fg$ and so~\eqref{eq:optStrategyGen} can be stated as
  \[
     \hpi = \mathop{\argmax}_{y\in \sC^0} \fg(y),
  \]
  where we note that $\sC^0$ is simply a subset of $\R$ because $F^R$ is deterministic and time-independent. In particular, the optimal trading strategy $\hpi$ is given by a constant.
  Moreover, setting $a:=\tfrac{p}{1-p}\max_{y\in\sC^0} \fg(y)$, the explicit formula for the opportunity process is
  \[
    L_t=a^{p-1}\big[(1+a)e^{a (T-t)} - 1\big]^{1-p}
  \]
  and then by~\eqref{eq:consumptionFeedback} the optimal propensity to consume is
  \[
    \hkappa_t=  1/L_t^{1/(1-p)}  = a\big[(1+a)e^{a (T-t)} - 1\big]^{-1}.
  \]
  We refer to~\cite[Theorem~3.2]{Nutz.09c} for the proof and references to related literature.
\end{Example}

\appendix

\section{Dynamic Programming}\label{ap:DynamicProg}

This appendix collects the facts about dynamic programming which are used in this paper.
Recall the standing assumption~\eqref{eq:PrimalProblemFinite}, the
set $\cA(\pi,c,t)$ from~\eqref{eq:admissibleAftert} and the process $J$ from~\eqref{eq:valueProcJ}.
We begin with the lattice property.

\begin{Fact}\label{fa:latticeProperty}
  Fix $(\pi,c)\in\cA$ and let $\Gamma_t(\tc):=E[\int_t^T U_s(\tc_s)\,\mu^\circ(ds)|\cF_t]$. The set $\{\Gamma_t(\tc):\, \tc\in\cA(\pi,c,t)\}$ is upward filtering for each $t\in[0,T]$.

  Indeed, if $(\pi^i,c^i)\in\cA(\pi,c,t)$, $i=1,2$, we have
  $
    \Gamma_t(c^1)\vee\Gamma_t(c^2)=\Gamma_t(c^3)
  $
  for $(\pi^3,c^3):=(\pi^1,c^1)1_A+(\pi^2,c^2)1_{A^c}$ with $A:=\{\Gamma_t(c^1)>\Gamma_t(c^2)\}$.
  Clearly $(\pi^3,c^3)\in\cA(\pi,c,t)$. Regarding Remark~\ref{rk:constraintsDef}, we note
  that $\pi^3$ satisfies the constraints if $\pi^1$ and $\pi^2$ do.
\end{Fact}

\begin{Prop}\label{pr:martOptPrinciple}
  Let $(\pi,c)\in\cA$ and
  $
    I_t(\pi,c):=J_t(\pi,c)+\int_0^t U_s(c_s)\,\mu(ds).
  $
  If $E[\,|I_t(\pi,c)|\,]<\infty$ for each $t$, then $I(\pi,c)$ is a supermartingale having a c\`adl\`ag version.
  It is a martingale if and only if $(\pi,c)$ is optimal.
\end{Prop}

\begin{proof}
  The technique of proof is well known; see El~Karoui and Quenez~\cite{ElKarouiQuenez.95} or Laurent and Pham~\cite{LaurentPham.99} for arguments in different contexts.

  We fix $(\pi,c)\in\cA$ as well as $0\leq t\leq u\leq T$ and prove the supermartingale property.
  Note that $I_t(\pi,c)=\esssup_{\tc\in\cA(\pi,c,t)} \Upsilon_t(\tc)$ for the martingale
  $\Upsilon_t(\tc):=E\big[\int_0^T U_s(\tc_s)\,\mu^\circ(ds)\big|\cF_t\big]$. (More precisely, the expectation is well defined with values in $\R\cup\{-\infty\}$ by~\eqref{eq:PrimalProblemFinite}.)

  As $\Upsilon_u(\tc)=\Gamma_u(\tc)+\int_0^u U_s(\tc_s)\,\mu(ds)$, Fact~\ref{fa:latticeProperty} implies that there exists a sequence $(c^n)$ in $\cA(\pi,c,u)$
  such that $\lim_n \Upsilon_u(c^n)=I_u(\pi,c)$ $P$-a.s., where the limit is monotone increasing in $n$. We conclude that
    \begin{align*}
     E[I_u(\pi,c)|\cF_t]
      &= E[\lim_n \Upsilon_u(c^n)|\cF_t]
       = \lim_n E[\Upsilon_u(c^n)|\cF_t] \\
      &\leq  \esssup_{\tc\in\cA(\pi,c,u)} E[\Upsilon_u(\tc)|\cF_t]
       = \esssup_{\tc\in\cA(\pi,c,u)} \Upsilon_t(\tc) \\
      &\leq \esssup_{\tc\in\cA(\pi,c,t)} \Upsilon_t(\tc)
        = I_t(\pi,c).
  \end{align*}

  To construct the c\`adl\`ag version, denote by $I'$ the process obtained by taking the right limits of $t\mapsto I_t(\pi,c)=:I_t$ through  the rational numbers, with $I'_T:=I_T$.
  Since $I$ is a supermartingale and the filtration satisfies the ``usual assumptions'', these limits exist $P$-a.s.,
  $I'$ is a (c\`adl\`ag) supermartingale, and $I'_t\leq I_t$ $P$-a.s.\ (see Dellacherie and Meyer~\cite[IV.1.2]{DellacherieMeyer.82}). But in fact, equality holds here because for all $(\tpi,\tc)\in\cA(\pi,c,t)$ we have
  \[
    \Upsilon_t(\tc)= E\Big[\int_0^T U_s(\tc_s)\,d\mu^\circ\Big|\cF_t\Big]=E[I_T(\tpi,\tc)|\cF_t]=E[I_T|\cF_t]\leq I'_t
  \]
  due to $I_T=I'_T$, and hence also $I'_t\geq \esssup_{\tc\in\cA(\pi,c,t)} \Upsilon_t(\tc)=I_t$.
  Therefore $I'$ is a c\`adl\`ag version of $I$.

  We turn to the martingale property. Let $(\pi,c)$ be optimal,
  then $I_0(\pi,c)=\Upsilon_0(\pi,c)=E[I_T(\pi,c)]$, so the supermartingale $I(\pi,c)$ is a martingale. Conversely, this relation states that $(\pi,c)$ is optimal, by definition of $I(\pi,c)$.
\end{proof}

The following property was used in the body of the text.

\begin{Fact}\label{fa:stabilityProperty}
  Consider $(\pi,c),(\pi',c')\in\cA$ with corresponding wealths $X_t, X_t'$ at time $t\in [0,T]$ and $(\pi'',c'')\in\cA(\pi',c',t)$. Then
  \[
    c1_{[0,t]} + \frac{X_t}{X_t'}\,c''\, 1_{(t,T]} \;\in \cA(\pi,c,t).
  \]
  Indeed, for the trading strategy $\pi1_{[0,t]}+\pi''1_{(t,T]}$, the corresponding wealth process is
  $X1_{[0,t]}+\frac{X_t}{X_t'} X''1_{(t,T]}>0$ by~\eqref{eq:wealthDefinition}.
\end{Fact}

\section{Martingale Property of the Optimal Processes}
The purpose of this appendix is to provide a statement which follows from~\cite{KaratzasZitkovic.03} and is known to its authors, but which we could not find in the literature. For the case without intermediate consumption, the following assertion is contained in~\cite[Theorem~2.2]{KramkovSchachermayer.99}.

\begin{Lemma}\label{le:MartingaleOfDuality}
  Assume~\eqref{eq:BoundsR} and~\eqref{eq:ELMMexists}. Let $(\pi,c)\in\cA$, $X=X(\pi,c)$ and $Y\in\sY^{\sD}$, then
  \[
   Z_t:=X_tY_t+\int_0^t c_sY_s\,\mu(ds),\quad t\in[0,T]
  \]
  is a supermartingale. If $(X,c,Y)=(\hX,\hc,\hY)$ are the optimal processes solving the primal and the dual problem, respectively,
  then $Z$ is a martingale.
\end{Lemma}

\begin{proof}
  It follows from~\cite[Theorem~3.10(vi)]{KaratzasZitkovic.03} that $E[Z_T]=E[Z_0]$ for the optimal processes, so it suffices to prove
  the first part.

  (i)~~Assume first that $Y\in\sY^{\sM}$, i.e., $Y/Y_0$ is the density process of a measure $Q\approx P$.
  As $\sY^{\sM}\subseteq \sY^*$, the process $X+\int c_u\,\mu(du)=x_0+\int X_- \pi\,dR$ is a $Q$-supermartingale, that is,
  $E^Q[X_t+\int_0^tc_u\,\mu(du)|\cF_s]\leq X_s+\int_0^s c_u\,\mu(du)$ for $s\leq t$. We obtain the claim by Bayes' rule,
  \[
    E\Big[X_tY_t+\int_s^t c_uY_u\,\mu(du)\Big|\cF_s\Big]\leq X_sY_s.
  \]

  (ii)~~Let $Y\in\sY^{\sD}$ be arbitrary, then by~\cite[Corollary~2.11]{KaratzasZitkovic.03} there is a sequence $Y^n\in\sY^{\sM}$ which Fatou-converges to $Y$. Consider the supermartingale $Y':=\liminf_n Y^n$. By \v{Z}itkovi\'c~\cite[Lemma~8]{Zitkovic.02}, $Y'_t=Y_t$ $P$-a.s.\ for all $t$ in a (dense) subset $\Lambda\subseteq[0,T]$ which contains $T$ and whose complement is countable. It follows from Fatou's lemma and step (i) that $Z$ is a supermartingale on $\Lambda$; indeed, for $s\leq t$ in $\Lambda$,
  \begin{align*}
    E\Big[X_tY_t+\int_s^t c_uY_u\,\mu(du)\Big|\cF_s\Big]
    & =  E\Big[X_tY'_t+\int_s^t c_uY'_u\,\mu(du)\Big|\cF_s\Big] \\
    & \leq \liminf _n E\Big[X_t Y^n_t+\int_s^t c_u Y^n_u\,\mu(du)\Big|\cF_s\Big]\\
    & \leq \liminf_n X_s Y^n_s = X_sY_s\quad P\mbox{-a.s.}
  \end{align*}
  We can extend $Z|_{\Lambda}$ to $[0,T]$ by taking right limits in $\Lambda$ and obtain a right-continuous supermartingale $Z'$ on $[0,T]$, by
  right-continuity of the filtration.
  But $Z'$ is indistinguishable from $Z$ because $Z$ is also right-continuous. Hence $Z$ is a supermartingale as claimed.
\end{proof}

\bibliography{C:/Users/numadmin/Documents/tex/stochfin}
\bibliographystyle{plain}

\end{document}